# Thermostat Interactions and Human Discomfort: Uncovering Spatiotemporal Variabilities in a Longitudinal Study of Residential Buildings

**SungKu Kang**[1,2+], **Maharshi Pathak**[2+], **Kunind Sharma**[2+], **Emily Casavant**[2], **Katherine Bassett**[2], **Misha Pavel**[3], **David Fannon**[2,4], **Michael Kane**[2,*]

[1] Korea Aerospace University, Department of Mechanical and Aircraft System Engineering, Goyang-si, Gyeonggi-do, 10540, Republic of Korea
[2] Northeastern University, Civil and Environmental Engineering Department, Boston, MA 02115 United States
[3] Northeastern University, Khoury College of Computer Sciences, Boston, MA 02215, United States
[4] Northeastern University, School of Architecture, Boston, MA 02215, United States
* mi.kane@northeastern.edu
+ these authors contributed equally to this work and are listed alphabetically

## ABSTRACT

Growing variable renewable energy and electrification of heating and transportation are intensifying the challenge of operating the electric grid. However, current demand response (DR) approaches compromise their efficacy by neglecting human-building interactions (HBIs). For example, utilities may increase thermostat setpoints on the hottest days of the year, reducing the strain on the grid but making occupants uncomfortable and frustrated. To better understand HBIs in residential buildings, 41 people in 20 homes in two climates participated in a 6-month study. Timestamps from app-based thermal comfort surveys and thermostat interactions were synchronized to time-series building systems data, resulting in the largest-of-its-kind HBI dataset. These survey data are compared to predictions from industry-standard thermal comfort models. Statistically significant results show that these models, developed under steady-state assumptions, yield greater error magnitudes and/or biases when spatiotemporal temperature variations exceed 2°F. The mean spatial variation within homes in the dataset was 4°F, thermostat DR control would commonly exacerbate such temporal variation. The results highlight opportunities for improving DR load-control algorithms through a paradigm shift to modeling discomfort rather than comfort, increasing the use of low-cost sensors, and incorporating dynamic models of occupant behavior.

## Introduction

Modernization of the U.S. electric grid aims to enhance resilience, efficiency, and reliability while shifting towards renewable energy. These goals are hampered by the uncontrollable nature of wind and solar energy[1], increasing demand from the electrification of heating and transportation, and increasing weather extremes[2,3]. New paradigms for demand-side management promise to overcome these challenges.

Buildings account for 75% of total electricity consumption. Controlling this load could help balance uncontrollable renewables and reduce peak demands thus offsetting the need for costly infrastructure improvements. These strategies fall under the umbrella of Grid-interactive Efficient Buildings (GEBs), in which a combination of HVAC systems, other electric loads in buildings, on-site generation, or energy storage can shed, shift, or modulate the building's electrical energy consumption[4]. The Department of Energy (DoE) estimates that by 2030 the potential annual benefit of GEBs to the U.S. electric grid could amount to between $8 billion to $18 billion, while reducing total power sector $CO_2$ emissions by 6%[5,6]. In the U.S., GEBs present peak-demand reduction opportunities of 6.3 – 17.4 GW from current 7.9 GW levels and annual energy savings of 166 – 622 TWh by 2030. However, evaluations of these demand flexibility solutions based on economic, technological, and environmental metrics[7] often overlook effects on building occupants and the quality of services provided by the electric grid, such as comfort provided by HVAC systems.

Demand response (DR) programs, the currently prevalent deployed GEB technology, aim to shed or shift buildings' peak demand, most often during the summer peak cooling season by adjusting thermostat setpoints or simply switching-off systems like air conditioning. DR may be applied to any electric loads such as water heaters, EV chargers, pool pumps, and industrial machinery; however, HVAC DR is common in residential settings and is the primary concern of this study. DR strategies may be classified based on the role of customers. In automated DR strategies (also known as active, dispatchable, or direct load control), the DR provider (often the electric utility) adjusts the customers' thermostat setpoint or turns off their HVAC system for a fixed period via the internet or a radio signal to reduce peak electricity demand on the grid. More advanced active DR strategies may try to shift demand earlier in the day by pre-cooling during off-peak hours. In nearly all cases, customers can override the changes made by the DR provider; however, the customer may forgo DR financial incentives when opting-out.[8] One study of an active DR program found up to 30% of customers overring controls during DR events[9]. Customers have an even greater influence on *indirect* DR (also known as traditional DR, non-dispatchable, or passive), in which DR provider calls, texts, or otherwise tries to convince customers to reduce their energy consumption during peak periods. Participant behavior in indirect DR programs can be challenging to predict and unreliable[10].

In both strategies, the DR programs' failure to provide load reductions promised to grid operators decreases grid reliability, increases emissions, and results in millions of dollars of penalties.[11] DR providers often have limited aggregate information about their performance (e.g., MW of load shed, time of day, weather), which they use to create statistical models predicting the amount of load they can shed during future events. These models do not explicitly account for occupant comfort and associated human-building interactions (HBI). However, the increasing use of smart thermostats can offer new insights for DR providers such as setpoints, indoor temperatures, occupancy, and history.

Early research on thermal comfort in the built environment yielded standard models such as Fanger's predicted mean vote (PMV) and percent population dissatisfied (PPD), which were codified into nearly-ubiquitous standards for building design and operation, such as ASHRAE 55 *Thermal Environmental Conditions for Human Occupancy.*[12] However, these models were based on the heat balance of humans in controlled climate chamber experiments, and for air-conditioned environments under steady-state conditions[12,13] and so may not represent residential settings[14] nor reflect the temperature transience and variations found in homes during DR events.[15] Other known limitations of the PMV model include insensitivity to measurement error; oversimplification of the heat balance model; failure to assess adaptive opportunities (e.g., behavior changes and environmental control); and neglecting contextual factors (e.g., climate, building type, season, and age).[16] For decades, there has been a wide variety of subsequent research and assessment through field studies across building types, as well as studies focused on the effects of natural ventilation, on personalized comfort systems, and the influence of personal and environmental variables on thermal comfort[17]. This exponential increase in research since the 1970s[17] also enabled
the development of alternatives to overcome specific challenges and improve predictive power. Newer approaches like the Advanced Human Thermal Comfort Model[18], and subsequent Advanced Comfort Tool[19] based on a human physiological model,[20] models which account for transient and non-uniform environments,[21] and the widely adopted Adaptive Thermal Comfort Model considering prevailing outdoor air temperature[12,22] help engineers design HVAC systems to provide comfortable indoor environments. Yet occupants in residential settings continue to report discomfort in ways that are not predicted by standard models[14,23–25]. Further, models of human thermal comfort present a particular challenge for DR operators who wish to predict whether and when building occupants may participate in an indirect DR event or override an automated one. Existing engineering models rely on personal and environmental factors such as metabolism, clothing, indoor radiant environment, air velocity, and humidity[12,13] which can be impractical for DR providers to accurately estimate.[16,26] New psychophysiological models like social cognitive theory[27], game theory[28], and perceptual control theory[29] have been proposed as future research directions to close the comfort-to-behavior knowledge gap but to date account only for steady-state, rather than dynamic human thermal adaptations.

This study evaluates methods for predicting occupant comfort in residential GEBs, identifies approaches for GEB service providers to improve reliability and customer satisfaction, and defines open questions for the research community. Three key aspects of this study drive its significance:

1. an analysis of the largest-of-its-kind residential HBI dataset of timestamped thermal comfort surveys and thermostat interactions synchronized with contextual building systems data;
2. a GEB-focused critique of industry-standard thermal comfort models that illuminates opportunities for improvement that would not be apparent with common critiques that rely on steady-state assumptions; and
3. time-series sensor data that enables an analysis of HBI dynamics that would not be possible with traditional comfort survey data and environmental data at that point-in-time.

The next section describes the quantitative and qualitative data collection methods employed in 20 homes and with their occupants over 6-months, along with methods for analyzing the data to understand occupant thermal comfort and behavior. The Results and Discussion sections present three key observations from this analysis:

1. an evaluation of the ASHRAE 55 PMV model and the adaptive comfort model under temporal indoor temperature variations that may be experienced in GEBs;
2. the prevalence of spatial indoor temperature variations and their impact on comfort predictions; and
3. patterns in occupants' manual thermostat setpoint changes.

Finally, a conclusion of the findings and a roadmap for future research completes the manuscript.

## Methods

### The Whole Energy Homes (WEH) Project

The Whole Energy Home (WEH) project, funded by the US Department of Energy (Award Number DE-EE0009154), aims to investigate interactions between occupants, HVAC systems, homes, and the grid with the long-term goal of realizing the potential of GEBs[30]. This manuscript analyzes data generated for 6-months (spring through autumn) of the first phase during which the research team monitored building performance and occupant behavior without intervention. During the second phase, the team will intervene, simulating DR events to measure the responses of the buildings and occupants to these active HVAC controls. At the end of the project, the anonymized dataset will be publicly available to provide the foundation for developing predictive models of occupant behavior and controls for DR programs.

The WEH project is ongoing and expanding. The analysis herein considers 41 participants in 20 homes in two different climates: Massachusetts and Colorado. All participants provided informed consent, and this human-subjects research was conducted in compliance with applicable guidelines and regulations, following a protocol approved by the Northeastern University Institutional Review Board (IRB #21-07-01). Participants were recruited from the greater Boston, MA and Denver, CO areas through social media platforms and community organizations. All participants live in owner-occupied single-family detached homes with central forced-air heating and air conditioning systems compatible with internet-connected smart thermostats. Enrollment of participants occurred from December 2021 through April 2022, and included four steps: (1) collecting qualitative data on occupant socio-economic demographics, environmental attitudes, and other psychological factors through individual participant interviews and surveys; (2) performing an energy assessment on each home; (3) installing a sensing and control suite for long-term monitoring; and (4) installing an app on the phones of adult occupants' in the homes enabling them to adjust the thermostat setpoints and respond to just-in-time micro-surveys known as ecological momentary assessments (EMA) regarding occupant thermal comfort and perception. The subsections below describe the digital infrastructure for supervisory control and data acquisition; systems and methods for home environmental monitoring and control; and occupant interactions and assessment.

*Digital Infrastructure of Supervisory Control and Data Acquisition*
To integrate data from disparate sources and provide a platform for customizable automation and interfaces, the research team deployed an open-source home automation software Home Assistant (HA) [31] server for each home, on computing infrastructure controlled by the team. HA connects the thermostats, smart plugs, weather data, and apps on participants' phones through web APIs. Data are either sampled when values change or at 5-minute intervals and recorded in an SQL database. All communication occurs through secure TLS web connections, and HA provides various levels of user access controls to provide appropriately granular access to research administrators, and analysts, and a simplified user-friendly interface for the participants themselves.

*Systems and Methods for Home Environmental Monitoring and Control*
Quantitative data on the home environment, HVAC system, and occupants were collected by smart thermostats, passive infrared (PIR) occupancy and temperature sensors, window/door sensors, and smart plugs installed by the research team as illustrated in **Figure 1**. Smart thermostats were installed in each HVAC zone (see Supplementary Table 1 in Supplemental Information for the number of zones in each home) that provide temperature, humidity, and occupancy in the room with the thermostat. The thermostat also reports to HA the setpoint (which HA and occupants adjust through the app), and runtimes for each HVAC component (e.g., air conditioning compressor was running along with circulation fan). Occupancy and temperature sensors were placed throughout each home (typically about six), as well as window/door sensors to monitor when occupants increase natural ventilation (typically about six). These devices report to HA by communicating through the thermostat. Smart plugs were installed on portable devices that may be used to modify thermal comfort such as electric heaters or window air conditioners. Weather conditions such as minute-scale temperature, humidity, wind, cloud cover, and air pressure are obtained from Internet weather services.

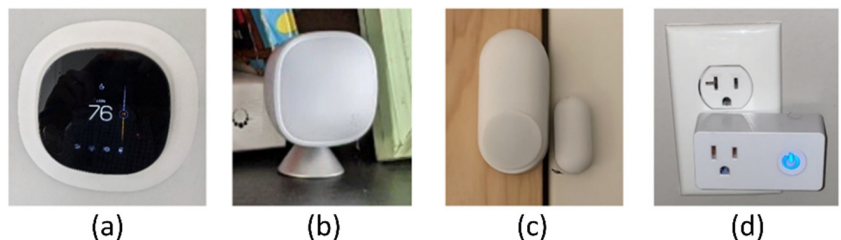

(a)     (b)     (c)     (d)

**Figure 1.** Smart sensors used in WEH project: (a) smart thermostat, (b) occupancy and temperature sensor, (c) window/door sensor, and (d) smart plug.

*Occupant Interactions and Assessment*
Data from the sensor suite described above are commonly found in datasets such as the ecobee Donate Your Data (DYD) dataset[32] or the Pecan Street dataset[33]. However, the WEH study goes beyond these measurements seeking *granular qualitative* and *quantitative* data on individual occupants' presence, perception, mental state, and behavior. To this end, three approaches were integrated, all implemented in the HA app on the participants' phones: tracking occupants' **location**; tracking occupants' **interactions** with the thermostat; and conducting **just-in-time EMAs**. To preserve privacy, participants' location was geo-coded only as 'home' or 'away.' Participants were asked to interact with the thermostat (i.e., adjust setpoints) through the app (**Figure 2**(a)) instead of the physical thermostat so that the data can illuminate who in the home is adjusting the thermostat. To understand qualitative personal factors that affect HBI, the technique of EMA—in which push notifications are sent to the participants to complete a brief survey—was borrowed from the health sciences. As outlined by an Annual Review of Clinical Psychology report[34], the term 'ecological' refers to the in-situ nature of the sampling strategy which allows conclusions drawn from survey responses to be more informative of real-world settings than data collected in controlled lab settings. The term 'momentary' refers to the in-real-time nature of the sampling strategy, where questionnaires only ask participants about their current state, such as their instantaneous observations, perceptions, or feelings. This strategy minimizes the questionnaire's reliance on memory recall, which is often systematically biased [34] . Repeated survey sampling over time enables EMA to capture changes in the psychological state and situational observations of participants over time.

Building scientists increasingly employ EMA as a method to supplement post-occupancy evaluations, assess indoor environmental quality (IEQ), and study occupant thermal comfort[35,36], including longitudinal studies[37]. Some studies, including the present work, pair EMA sampling with simultaneous IEQ sensing, not only enabling tracking of the occupant state over time but also in relation to other dimensions such as indoor air temperature. EMA additionally allows for advanced and customized sampling methods to ensure survey responses are captured over the whole 'range of interest' for a given research objective[35].

EMA notifications were sent to participants at random times during waking hours when they were home. Then the participants could complete the EMA anytime thereafter, with the time of the EMA recorded as when they submitted their response, not when the notifications were sent. Participants may also dismiss the notification and not complete the EMA. The EMA consists of three questions on thermal sensation, satisfaction, and preference as shown in **Figure 2**(b-d) respectively.

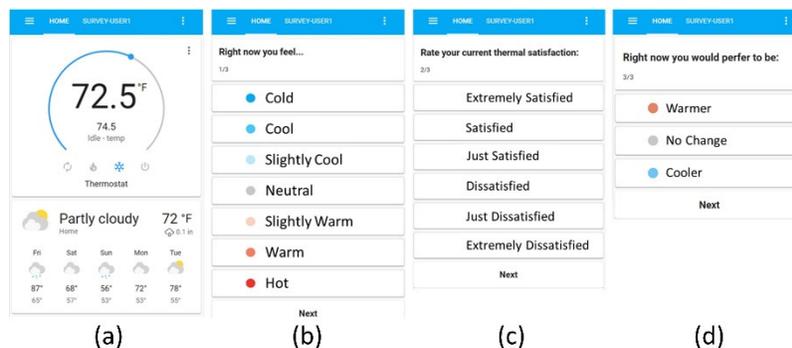

(a)          (b)          (c)          (d)

**Figure 2.** Home Assistant (HA) interface seen by WEH participants. (a) The thermostat setpoint can be adjusted by the semi-circular blue slider and seen in the small print. The large print shows the current temperature measured by the thermostat. When the participant clicks the push notification requesting them to complete an EMA, they see questions (b-d), collecting data on their thermal sensation, satisfaction, and preference.

*Unique Features of the Dataset*
The WEH dataset incorporating the components above offers several features not provided by datasets founded solely on sensor data. First, the dataset combines two temporally registered data series: (1) building system data collected from smart sensing infrastructure and (2) EMA responses collected from the occupants. The latter serves as the ground truth of thermal sensation and satisfaction when evaluating thermal comfort. In addition, the dataset annotates sensor and occupancy data with rich, human-readable spatial context that can be associated with EMA responses. Rather than identify sensors by letters or numbers, the WEH data employs descriptive tags like `basement living room' and 'upstairs playroom'. Together these features result in a unique dataset, enabling detailed analyses of HBI and its implications for the design and control of GEBs based on ground truth thermal sensation and satisfaction; the progression of occupancy; spatial temperature variations, and their influence on occupant comfort. Important considerations for the design of controls for GEBs.

**Evaluation of Comfort and Behavior in Longitudinal Data from WEH Phase 1**
As presented in the introduction, current DR programs and future GEBs lack robust engineering models of occupant thermal comfort behavior suited to design grid-interactive thermostat control algorithms (e.g., thermostat setback and duration). This manuscript's analysis of data from Phase 1 of the WEH project tests the hypotheses that spatiotemporal temperature variations play a significant role in occupant comfort and behavior and aims to better understand when occupants override or manually change thermostat setpoints.

The following three subsections describe the (1) data curation; (2) assumptions and features used in evaluating the ASHRAE 55 PMV comfort model and adaptive comfort model under steady-state and transient conditions; and (3) assumption and features defined for the evaluation of manual setpoint changes (MSCs).

*Data curation*
Temperature, occupancy, thermostat setpoints/interactions, and EMA responses from April through September 2022 for nineteen of twenty homes were curated, except for one home that did not provide EMA responses. Data were segmented to remove gaps, and continuous variables such as temperature and occupancy were converted from event-based records to 5-minute interval data to facilitate analysis.

*Standard comfort models' features and assumptions*
To evaluate standard methods for predicting occupant thermal comfort, the ASHRAE 55 PMV[12] and adaptive thermal comfort models[12,22] are compared against participants' EMA responses. The WEH study was not able to collect all the types of data required by these models, as such the following reasonable and common assumptions are made. The indoor **mean radiant temperature** is assumed to be equal to the ambient air temperature at the thermostat and remote temperature sensors. Humidity was only measured at the thermostat, not at remote temperature sensors, so **humidity** was assumed to be equal throughout the home. The **constant relative airspeed** of 0.3 ft/s is widely accepted to represent standard low indoor air movement conditions based on the SET model[12]. The **prevailing outdoor temperature** for the adaptive thermal comfort model was calculated for a 7-day running average reported from ecobee, using the pythermal comfort[38] package in Python. Considering the data was collected in the cooling and shoulder season, a constant **clothing insulation** value of 0.51 Clo is considered representative[12]. A **metabolic rate** of 1.1 Met, consistent with "seated typing" was assumed for all occupants[38].

The thermal sensation and acceptability of the PMV and adaptive thermal comfort models are calculated based on the environmental conditions at the time the EMA response was submitted. The EMA questionnaires only asked about thermal sensation, satisfaction, and preference. Therefore, to evaluate the predictive power of the adaptive comfort model, responses indicating "just satisfied", "satisfied", or "extremely satisfied" were considered as acceptable thermal comfort. Similarly, "just dissatisfied", "dissatisfied", and "extremely dissatisfied" were considered as unacceptable conditions.

In traditional HVAC system design, the thermostat setpoint is assumed to be constant, and temporal temperature variations caused by the system cycling within the thermostatic deadband are assumed to be short. For longer and/or monotonic changes—called *drifts* and *ramps*—ASHRAE 55 defines time and temperature limits shown in **Table 1**. The assumption of constant setpoints would not be valid for automated DR of HVAC systems, thus greater temperature variations may be observed in GEBs, creating situations in which steady-state comfort models may not apply. To understand the temporal effect of temperature changes on the predictive power of comfort models, EMA responses for which the temperature at the thermostat changed more than $2.0°F$ in the previous 15-minutes are defined as "transient"; otherwise, the occupant is assumed to be in a steady-state thermal environment.

| Time period, hour | 0.25 | 0.5 | 1 | 2 | 4 |
|---|---|---|---|---|---|
| **Maximum Operative Temperature $t_o$ Change Allowed, °F** | 2.0 | 3.0 | 4.0 | 5.0 | 6.0 |

**Table 1.** Limits on temperature changes (drifts and ramps) over time, specified in ASHRAE 55[12].

*Thermostat override features and assumptions*

Understanding the behavioral dynamics that cause occupants to interact with the thermostat due to discomfort with their thermal environment is a central focus of our analysis. To effectively study these dynamics, it is beneficial to anchor our observations of behavior (i.e., MSCs) to a known initial condition, from which the MSC is time-referenced. Ideally, both the internal state (e.g., occupant comfort) and any external disturbances (e.g., changes in indoor temperature) would be known and initiated at this reference point. However, defining such a precise initial condition is challenging in practice.

In dynamical systems analysis, the internal state should be well-defined and observable. In the context of occupant thermal comfort, the 'state' of behavioral dynamics is ambiguous and can only be observed qualitatively. If an occupant reports feeling 'comfortable,' it can be interpreted as having no internal pressure to adjust the thermostat—a 'zero initial state' in terms of behavioral dynamics.

We consider MSCs that occur within 150 minutes of a defined reference point; as this window is greater than the time constants of behavior dynamics we previously observed[39]. We established two types of reference points to serve as known initial conditions for our analysis. The first type occurs when a participant submits an EMA response indicating a 'neutral' thermal sensation. This response suggests that the occupant is comfortable at that moment. However, when an MSC is observed within 150 minutes of a 'neutral' EMA response, it is unclear when the external disturbance that initiated the behavioral dynamics began. We assume the disturbance occurs after the EMA, otherwise the occupant would not have reported feeling neutral. (However, we recognize the possibility that the occupant had not yet perceived it.) Therefore, the time from the reference point ('neutral' EMA response) to the MSC may overestimate the actual time from the unobserved disturbance to the MSC, but it is the best approximation available.

The second type of reference point is used in the common case where an MSCs does not occur within 150 minutes after an EMA. In these cases, if a prior setpoint change occurred within 150 minutes before an MSC, that initiating setpoint change is defined as the reference. This initiating setpoint change could be automated (e.g., due to a schedule) or another MSC. We assume that the occupant was comfortable before and up to the time of this reference setpoint change. The reference setpoint change acts as an external initiating event that activates both the building's thermal dynamics and the occupant's thermal comfort behavioral dynamics. It is possible that the occupant does not immediately notice the setpoint change and must wait for the indoor temperature to adjust before feeling discomfort.

In both scenarios, we recognize that the actual time from when the environmental change leads to an MSC is likely shorter, by an unknown amount, than the observed time to MSC (TtMSC). Therefore, TtMSC represents an upper limit on the time interval between an MSC and the engagement of motivating factors. To account for potential confounding factors, we exclude cases where the interval between the reference point and the MSC is larger than 150 minutes or smaller than 5 minutes, as these may indicate either unrelated events or quick successive MSCs due to a sudden change of mind.

From the MSCs that occur within 150 minutes of a 'neutral' EMA or a previous setpoint change, we extract the following features for analysis:

(1) The number of MSCs;
(2) Which participant made the MSCs, tracked through thermostat adjustments in their app;
(3) The time of day and the day of the week the MSCs occurred;
(4) The time elapsed from a reference point to the MSC (*TtMSC*); and
(5) The magnitude of the MSC in degrees (*DoMSC*), calculated as the difference between the MSC setpoint and the reference setpoint.

It's important to note that the applicability of our analysis to grid-interactive efficient buildings (GEBs) assumes that the manual setpoint changes observed in this study's first phase are motivated similarly to occupants overriding a demand response (DR) setback on the thermostat. However, this assumption may not hold true for all scenarios involving DR setbacks. To address this limitation, the future second phase of our study aims to include automated thermostat setpoint changes, akin to those occurring during a DR event, to evaluate the validity of this assumption.

## Results and Discussion

The analysis results are presented in five parts: (1) an assessment of the adaptive comfort model's ability to predict thermal acceptable conditions, and more importantly for DR programs, the ability to predict thermally unacceptable conditions which a participant may try to override; (2) a comparison of the PMV model predictions to survey results under time-varying temperatures, highlighting the limitations of these models in predicting occupant satisfaction during demand response scenarios; (3) a comparison of the PMV model predictions to survey results considering the effect of spatial temperature variations across the home, highlighting importance and approach for considering these variations in DR programs; (4) an exploration of patterns of occupant comfort among homes and occupants and an exploration of patterns in MSC timing and degree; and (5) a discussion of how these results could lead to improved DR strategies.

### The Adaptive Comfort Model: Thermal Satisfaction and Dissatisfaction

The ASHRAE adaptive comfort model[12] is a popular tool used to design environments that 80% or 90% of occupants will find thermally 'acceptable' given the prevailing outdoor temperature, $T_{out}$. This study is interested in assessing the fitness of this model to design GEB controls that occupants find acceptable, or more importantly for GEBs, that occupants do not find so unacceptable that the occupant would override the controls. The assessment began by binning the EMA thermal satisfaction results by $1.5\ °F$ of prevailing outdoor temperature. **Figure 3**(a) shows a boxplot of indoor temperatures ($T_{in}$) for each bin where participants reported being thermally 'satisfied.' In contrast, **Figure 3** (b) shows the distribution of indoor temperatures where participants reported being thermally 'dissatisfied.' Both plots display shaded regions indicating where 80% and 90% of the population would find the environment 'acceptable' according to the adaptive comfort model.

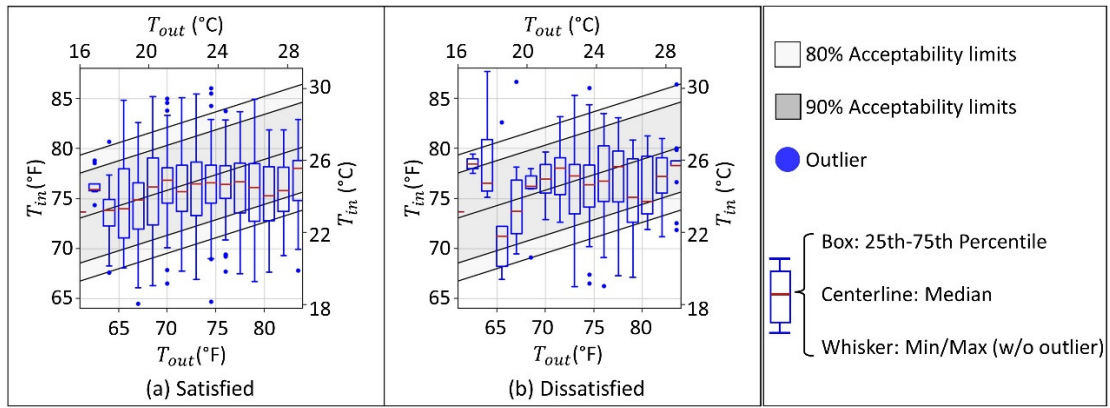

**Figure 3.** Distributions of thermal satisfaction from EMA responses for satisfied (a) and dissatisfied (b) responses, overlayed on the ASHRAE 55 adaptive thermal comfort model with the 80% and 90% acceptability limits.

The model accurately predicted occupants' thermal satisfaction, which is the model's intended purpose. Of the 'satisfied' responses (**Figure 3**(a)), 87.6% fall within the region where the model predicts that 80% of the population would find the environment 'acceptable.' However, **Figure 3** (b) shows that 78.7% of the 'dissatisfied' responses also fall within the 'acceptable' region. That is, the model is less effective at anticipating dissatisfaction, an essential aspect of designing DR controls.

**Table 2** establishes a confusion matrix to statistically assess the adaptive comfort model's ability to predict dissatisfaction. Here, a true positive (TP) indicates a *dissatisfied* EMA response in conditions *outside* the 80% acceptability limits, while *satisfied* responses *within* the limits are regarded as true negative (TN). Conversely, satisfied EMA responses *outside* the limit are regarded as false positive (FP), and dissatisfied responses within the limit are regarded as false negative (FN).

|  |  | **Predicted** | | **Total** |
|---|---|---|---|---|
|  |  | **Within 80% Acceptability Limit** (Negative) | **Outside 80% Acceptability Limit** (Positive) |  |
| **Actual** | **Satisfied Votes** (Negative) | 71.78% N=1,071 | 15.35% N=229 | 87.13% N=1,300 |
|  | **Dissatisfied Votes** (Positive) | 10.12% N=151 | 2.75% N=41 | 12.87% N = 192 |
| **Total Votes** |  | 81.90% N=1,222 | 18.10% N=270 | 100% $N_{total}$=1,492 |

**Table 2:** Confusion matrix for actual and predicted thermal satisfaction using the adaptive thermal comfort model in **Figure 4**.

The contrast between the high accuracy[1] (0.7453) and the low F1-score[2] (0.1775) indicates the model's low precision and sensitivity to predict *dis*satisfaction. Further, the small phi coefficient[3] (0.0325) indicates significant randomness in the prediction. This result suggests that the adaptive model works as intended when designing and operating occupant-controlled naturally ventilated spaces to maintain interior conditions that most of the population finds acceptable. However, the model is *not* suitable for DR providers who are most interested in predicting and minimizing instances when occupants find the indoor temperature *unacceptable* and might override the DR controls.

---

[1] (TP + TN)/(TP + TN + FP + FN), where TP: True positive, TN: True negative, FP: False positive, and FN: False negative.

[2] 2·TP/(2·TP + FP + FN)

[3] (TP·TN − FP·FN)/sqrt((TP + FP)·(FP + FN)·(TN + FP)·(TN + FN)), also known as Matthews correlation coefficient

**Temporal Temperature Variations and the PMV Model**
This section compares the actual thermal sensation vote (ASV) from each EMA response to the predicted mean vote (PMV) using environmental measurements from the thermostat and remote sensors. Thermal sensation votes range from -3 (hot) to +3 (cold) on a 7-point Likert scale. Thus, the PMV error (i.e., PMV minus ASV) theoretically ranges from -6 to +6, although the observed PMV error only ranges from -4 and +4. The boxplot in the center of **Figure 4** shows the PMV error distribution when the average temperature of sensors in occupied rooms ($T_{occu}$) drifted less than the ASRHAE 55[12] limits of ±2°F over the previous 15-minutes. These conditions should satisfy the thermodynamic steady-state conditions that the PMV model assumes. The point cloud of EMA response in **Figure 4** shows that most of the votes occurred under these steady-state temperature conditions ($|\Delta T_{occu,15min}| < 2°F$). This is consistent with homes that have few, if any, thermostat setpoint changes per day and are not providing DR services. The mean PMV error in the steady state region is nearly zero. However, the large standard deviation of 1.5 on the 7-point Likert scale where $|\Delta T_{occu,15min}| < 1°F$, indicates that there is little statistical significance between 'slightly warm' (-1), 'neutral' (0)', and 'slightly cool' (+1).

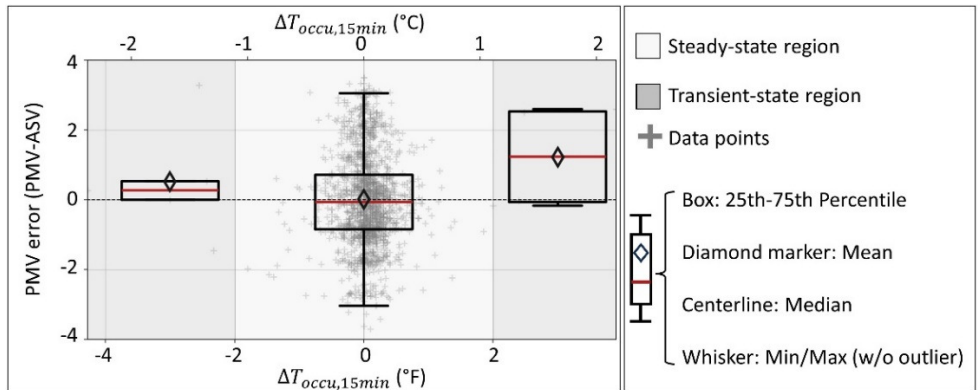

**Figure 4:** PMV error (i.e., PMV minus ASV) versus the average occupied room temperature drift in the 15-minutes before the EMA response ($\Delta T_{occu,15min}$). The region where $|\Delta T_{occu,15min}| < 2°F$ is considered steady state; otherwise, transient.

The non-zero mean PMV error in the left boxplot of **Figure 4** where the occupied temperature decreased more than 2°F, and the right side, where the temperature increased more than 2°F in the previous 15-minutes, suggests that the steady-state assumption of the PMV model may not hold in these transient regions. The following null hypothesis is tested using the Mann-Whitney U test: the PMV error is independent of steady-state or transient conditions. **Table 3** shows the test results: the probability of rejecting the null hypothesis is about 54% for the cooling temperature transient region and 89% for the warming transient region, respectively.

| Choice of X and Y | $U_x$ | $U_y$ | p-value |
|---|---|---|---|
| **X:** Data within steady-state region (group size: 1775) <br> **Y:** Data within transient region, negative $\Delta T_{occu,15min}$ (group size: 5) | 3603 | 5271 | 0.467 |
| **X:** Data within steady-state region (group size: 1775) <br> **Y:** Data within transient region, positive $\Delta T_{occu,15min}$ (group size: 4) | 1941 | 5158 | 0.117 |

**Table 3.** Mann–Whitney U test results from the PMV error values within temporally steady-state region ($|\Delta T_{occu,15min}| < 2°F$) and transient regions ($\Delta T_{occu,15min} \leq -2°F$ or $\Delta T_{occu,15min} \geq 2°F$) in **Figure 3**.

The relatively few samples in the transient regions and the large variation made it more difficult to show a statistically significant ($p < 0.05$) rejection of the null hypothesis. However, the result still suggests a potential correlation between downward drifting temperatures and PMV error during the cooling/shoulder season. These potentially biased responses during transients could guide future DR control design. For example, if a larger dataset of transient scenarios validates this finding, occupants reporting being warmer than predicted could allow for greater flexibility in pre-cooling.

**Spatial Temperature Variations and the PMV Model**

The WEH study quantified the spatial temperature variations within each home by collecting temperature and occupancy data from the thermostat and sensors in approximately five other rooms. The analysis compares the thermostat-measured temperature ($T_{tstat}$) to the temperature measured by sensors in rooms where occupancy is detected ($T_{occu}$). **Figure 5** plots the distribution of 'temperature variability' ($|T_{occu} - T_{tstat}|$) in each home. The left violin plots define variability as $\left(\text{avg}(|T_{occu}^i - T_{tstat}|)\right)$, the average absolute difference between measured temperatures $T_{occu}^i$ in occupied rooms $i$ and $T_{tstat}$. The right violin plots define variability as $\left(\max(|T_{occu}^i - T_{tstat}|)\right)$, the maximum absolute difference between $T_{occu}^i$ and $T_{tstat}$ at each 5-minute measurement from April through September 2022.

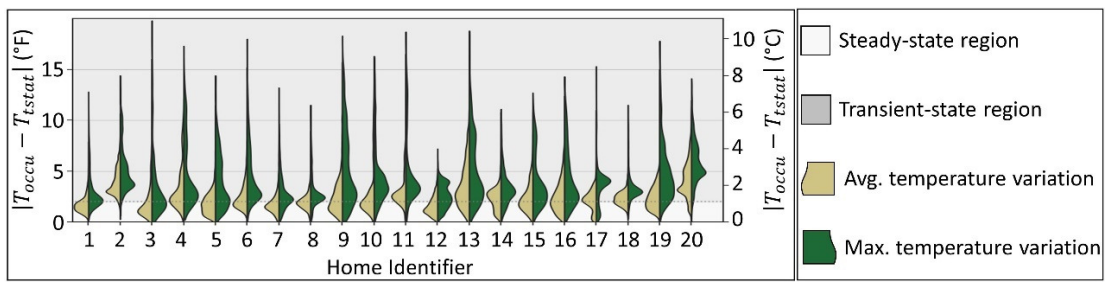

**Figure 5.** The distribution of average and maximum spatial temperature variations from field data. The horizontal line and shaded area above $2°F$ indicate that an occupant moving between spaces with such a temperature difference would experience a temperature drift outside the bounds specified by the ASHRAE 55 limits shown in **Table 1**. Detailed statistics of these data are provided in Supplementary Table 2 of the Supplemental Information.

The mean maximum spatial temperature variation across all 20 homes is $4.0°F$ with a standard deviation of $2.5°F$. While ASHRAE 55 does not specify spatial temperature variation limits, an occupant moving between spaces with a temperature difference greater than $2°F$ (as illustrated by the horizontal line in **Figure 5**), instantaneously experiences a temperature drift outside the ASHRAE 55 limits (**Table 1**). By this criterion, 13 homes have average spatial temperature variations that exceed $2°F$ at least 50% of the time.

Consider how five occupants experience these temperature variations in Home 2 located in Massachusetts. As shown in **Figure 6**, rooms that are regularly occupied (as indicated by the bubble size) have temperature differences from $T_{tstat}$ (as indicated by the bubble color) that regularly exceed $2°F$. These rooms are regularly cooler than $T_{tstat}$. However, Home 17 located in Colorado with three occupants, experiences occupied room temperature differences from $T_{tstat}$ that exceed $2°F$ cooler or $2°F$ warmer depending on the time of day.

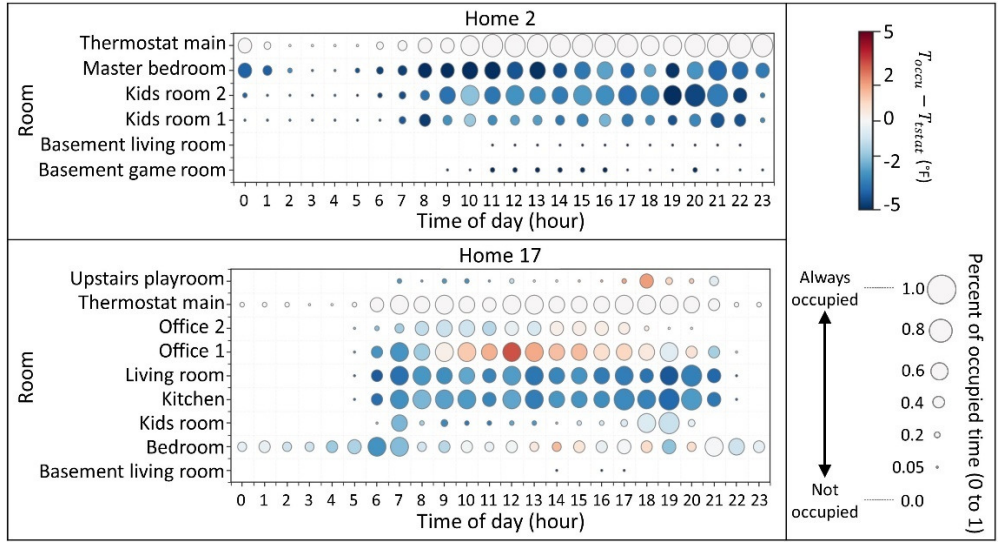

**Figure 6.** The progression of occupancy along the time of day and associated spatial temperature variation from the main thermostat temperature.

The following analysis aims to show the effect of these variations on PMV error. Here, the spatial variation is defined as the average temperature variation $\left(\text{avg}(T_{occu}^{i} - T_{tstat})\right)$. Considering that most residential thermostat installations only measure the indoor temperature at the thermostat, **Figure 7**(a) calculates the PMV using the temperature measured at the thermostat. Like the temporal PMV error analysis, the spatial PMV error is analyzed in three regions: spatially homogenous temperatures $|\text{avg}(T_{occu} - T_{tstat})| < 2°F$, average occupied room temperatures warmer than the thermostat $\text{avg}(T_{occu} - T_{tstat}) \leq -2°F$, and occupied temperatures cooler than the thermostat $\text{avg}(T_{occu} - T_{tstat}) \geq 2°F$.

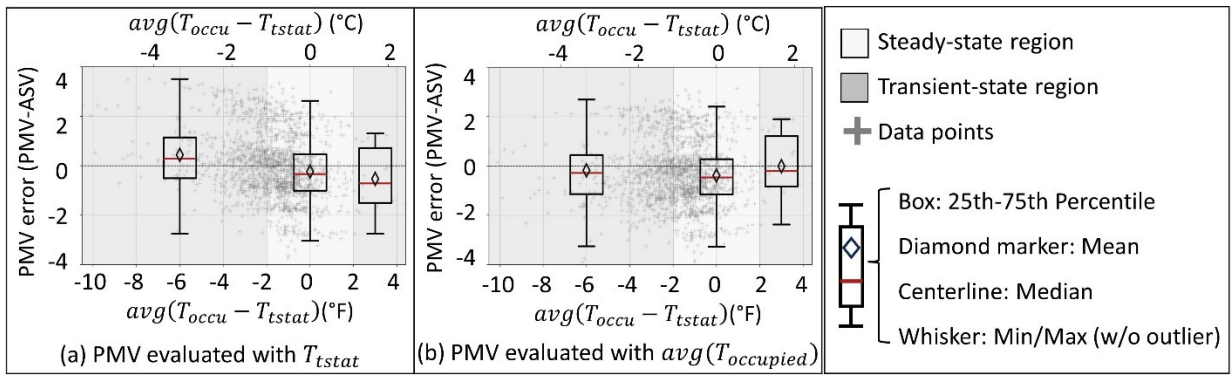

**Figure 7.** PMV error with respect to the actual thermal sensation vote (ASV) versus the average spatial temperature variation related to the thermostat ($\text{avg}(T_{occu} - T_{tstat})$), where PMV value is evaluated with a) $T_{tstat}$ and b) $avg(T_{occu})$. The region where $|\text{avg}(T_{occu} - T_{tstat})| < 2°F$ is considered spatially steady-state, otherwise transient.

The box plots of the PMV error in each region of **Figure 7**(a) show bias, even in the steady state region. Additionally, a downward trend in PMV error is apparent with increasing temperature variation. That is—unsurprisingly—when the home is cooler on average than the observed thermostat temperature, occupants tend to report being cooler than PMV would predict, and vice versa. To determine if this trend is significant, the following null hypothesis is tested using the Mann-Whitney U test: the PMV error bias is independent of spatially homogeneous or spatially varying conditions. **Table 4** shows the test results: the difference in PMV error is significant ($p = 5.23 \cdot 10^{-25}$) between the homogeneous and cooler conditions, but less significant ($p = 0.273$) between homogeneous and warmer conditions.

| Choice of X and Y | $U_x$ | $U_y$ | p-value |
|---|---|---|---|
| **X:** Data within steady-state region (group size: 1,079) <br> **Y:** Data within transient region, negative avg($T_{occu} - T_{tstat}$) (group size: 687) | 262,727 | 478,546 | 5.23e-25 |
| **X:** Data within steady-state region (group size: 1,079) <br> **Y:** Data within transient region, positive avg($T_{occu} - T_{tstat}$) (group size: 18) | 11,172 | 8,249 | 0.273 |

**Table 4.** Mann–Whitney U test results from the PMV error values within spatially steady-state region ($|avg(T_{occu} - T_{tstat})| < 2°F$) and transient regions ($avg(T_{occu} - T_{tstat}) \leq -2°F$ or $avg(T_{occu} - T_{tstat}) \geq 2°F$) shown in **Figure 7**(a), when PMV error is evaluated with $T_{tstat}$.

The analysis in **Figure 7**(b) aims to reduce PMV error by leveraging all available temperature and occupancy sensors. Specifically, the PMV calculation assumes the environmental temperature is the average measured temperature in occupied rooms avg($T_{occu}$) at the moment of the EMA response. Applying the same null hypothesis test yields (**Table 5**) a weaker but still significant ($p = 0.0044$) difference in PMV error between homogeneous and cooler conditions, and similar ($p = 0.249$) relationship between homogeneous and warmer conditions.

| Choice of X and Y | $U_x$ | $U_y$ | p-value |
|---|---|---|---|
| **X:** Data within steady-state region (group size: 1,079) <br> **Y:** Data within transient region, negative $avg(T_{occu} - T_{tstat})$ (group size: 687) | 340,871 | 400,401 | 0.0044 |
| **X:** Data within steady-state region (group size: 1,079) <br> **Y:** Data within transient region, positive $avg(T_{occu} - T_{tstat})$ (group size: 18) | 8,174 | 11,248 | 0.249 |

**Table 5.** Mann–Whitney U test results from the PMV error values within spatially steady-state region ($|avg(T_{occu} - T_{tstat})| < 2°F$) and transient regions ($avg(T_{occu} - T_{tstat}) \leq -2°F$ or $avg(T_{occu} - T_{tstat}) \geq 2°F$) shown in **Figure 7**, when PMV error is evaluated with $avg(T_{occu})$.

Comparing **Figure 7**(a) to **Figure 7**(b) and **Table 4** to **Table 5** illustrates the significant effect of leveraging a handful of additional temperature sensors in reducing PMV errors due to spatial temperature variations. **Table 6** further tests this significance with the null hypothesis that the PMV error is independent of whether $T_{tstat}$ or avg($T_{occu}$) is used to calculate the PMV. This hypothesis is rejected with statistical significance ($p = 9.89 \cdot 10^{-20}$ and $p = 0.0017$) where the occupied rooms are cooler than the thermostat and the homogeneous temperature regions respectively. This suggests that DR providers could improve their predictions of occupant (dis-)comfort by installing additional temperature sensors.

| Target region | $U_x$ | $U_y$ | p-value |
|---|---|---|---|
| Data within transient region, negative $avg(T_{occu} - T_{tstat})$ (group size: 687) | 302,833 | 169,136 | 9.89e-20 |
| Data within steady-state region (group size: 1,079) | 627,596 | 536,645 | 0.0017 |
| Data within transient region, positive $avg(T_{occu} - T_{tstat})$ (group size: 18) | 117 | 207 | 0.159 |

**Table 6.** Mann–Whitney U test results from **X:** the PMV error values evaluated with $T_{tstat}$ (**Figure 7** (a)) and **Y:** the PMV error values evaluated with $avg(T_{occu})$ (**Figure 7** (b)), for spatially steady-state and transient regions.

**Patterns of Thermostat Manual Setpoint Changes (MSCs)**

This analysis dives into thermostat usage patterns across different households and among occupants within the same household, providing additional insight into the population-level findings above. Each participant's thermal comfort temperature is estimated from the average measured temperature in all occupied rooms whenever an EMA response contains 'satisfied' or 'extremely satisfied' thermal satisfaction or 'no change' thermal preference. These thermal comfort distributions are shown in **Figure 8** for homes where both users submitted at least 30 EMAs. The average difference between an individual's mean comfort temperature and the population mean is $1.1°F$, while the average difference between users in the same household is only $0.65°F$. This could be beneficial for DR providers that can apply setbacks tuned to the preferences of occupants in each home.

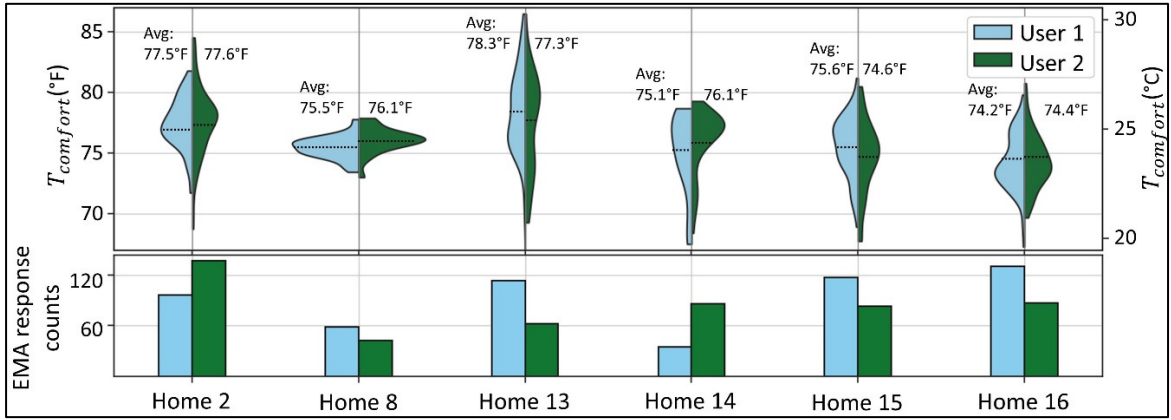

**Figure 8:** Comparison of comfort temperature distributions and EMA response counts between the occupants in each home.

Patterns in the number of MSCs, who is making MSCs, the MSCs time of day and week vs weekend, and *DoMSC* are observed in **Figure 9**. **Figure 10** shows these same data plus *TtMSC* in boxplots providing greater statistical insight. As in **Figure 8**, variations between homes and between users are apparent, but variations between homes are more evident than between users within a home. This could prove beneficial for DR providers who can create separate setbacks for each home but need not do so for each *occupant* within a home. **Figure 9** also shows the dark lines at 10PM, 9PM, and 10PM in Homes 1 (User 1), Home 16 (User 1), and Home 19 (User 2) respectively, which appear to show a consistent timing of MSCs in those homes. This behavior suggests a consistent override of schedule setpoint changes, perhaps due to the occupants' schedules (e.g., work or bedtime), and perhaps an opportunity to better align the thermostat settings with occupant needs. The mean *TtMSC* is often higher than its median for each home in **Figure 10**, indicating that MSCs occur more frequently shortly after a reference point, although a long tail, beyond two hours, exists. Homes 1 and 2 in **Figure 10** have non-overlapping *TtMSC* interquartile ranges, showing a statistically significant difference between the thermostat use behavior in these two homes, something for which a DR provider may wish to account.

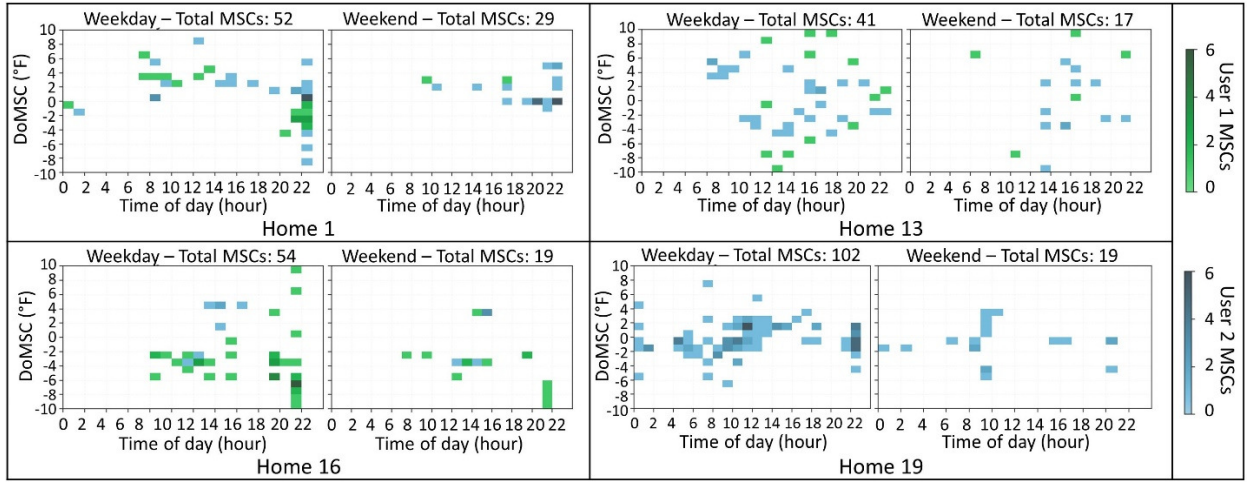

**Figure 9.** Visualization of the features related to manual setpoint changes (MSCs), showing different distributions across homes, users, days (weekday/weekend), and time of day.

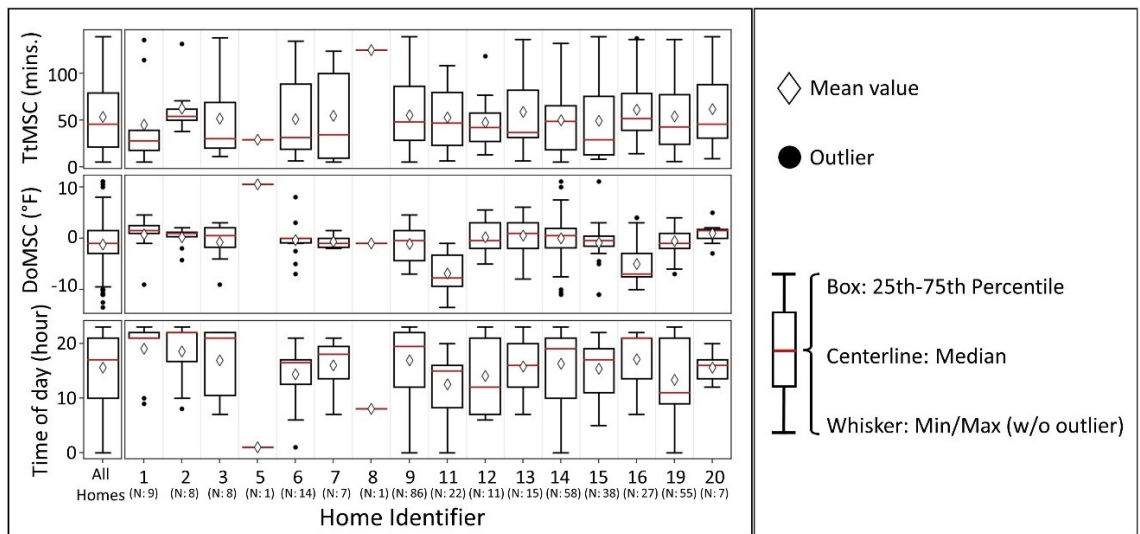

**Figure 10.** The comparison of the features related to manual setpoint changes (MSCs) between the entire population and each home.

The *TtMSC* varies significantly, even for each user. Some MSCs are not easily explained by an apparent schedule. A hypothesis for these observations is that users are overriding a recent setpoint change that made them uncomfortable. To investigate this hypothesis, the *TtMSC* and *DoMSC* for each EMA, along with MSC counts, are plotted in **Figure 11**. For a given DoMSC, the 50[th] percentile marker shows the percentage of MSCs, at the given DoMSC, that occurred before the TtMSC indicated by the marker. A line is then fit through these points to estimate the relationships between DoMSC and TtMSC.

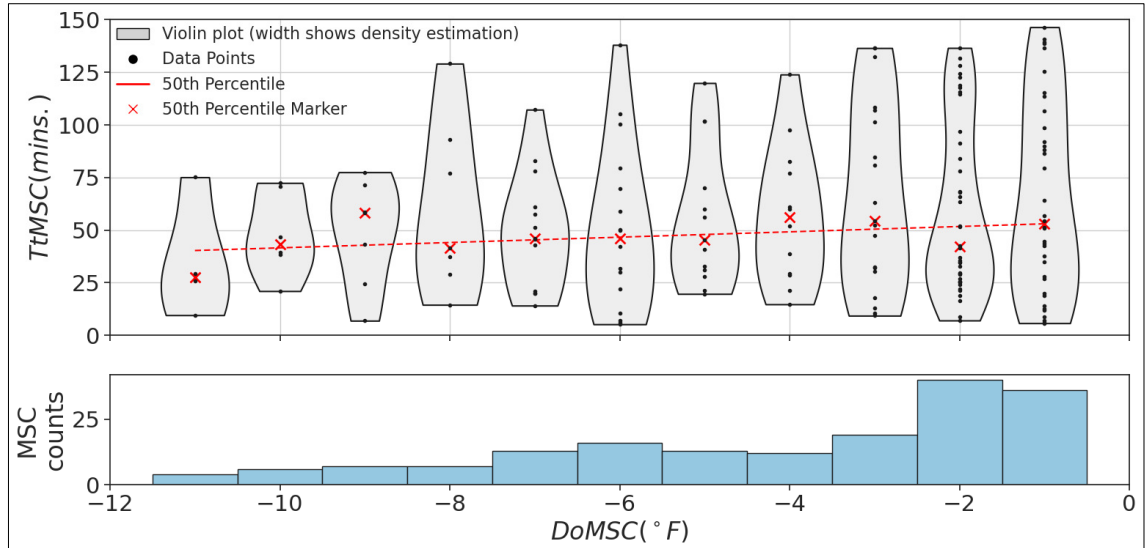

**Figure 11**: Time to Manual Setpoint Change (*TtMSC*) versus Degree of Manual Setpoint Change (*DoMSC*), associated with energy-intensive MSCs. The linear fit for the 50th percentile of the MSCs shows the negative correlation between the magnitude of *DoMSC* and *TtMSC*. The coefficients and RMSE of these linear fits are provided in Supplementary Table 3 of the Supplemental Information.

The slopes of the fitting curves suggest a negative correlation between the magnitude of *DoMSC* and *TtMSC* during the cooling season for 'energy-intensive' MSCs. This phenomenon, in which larger recent setpoint changes are overridden more quickly than smaller changes, may have a psychophysiological explanation in *thermal frustration,* defined as an accumulation of thermal discomfort over time, and when that accumulated frustration reaches a threshold the occupant takes action, for example overriding the thermostat. In this explanation, small amounts of discomfort will take longer to accumulate to the threshold than greater discomfort. A loose analogy in the context of DR could be made where 50% of participants will override a given setback amount (analogous to DoMSC) for the setback duration indicated by the line (analogous to TtMSC).

**Towards a Model-based Control Model of Occupant Thermostat Use Behavior for GEBs**
Given the limitations of current steady-state comfort models, we envision that the future dynamic comfort model to support residential DR programs should provide the following capabilities. First, the future dynamic model should incorporate **spatial and temporal temperature variations**, by using the rate of temporal temperature change. While this analysis used *average* occupied room temperature to indirectly incorporate spatial temperature variations, the occupancy and temperature of each room may not be available to DR providers due to the lack of sensors or data privacy concerns. To address the gap, an alternative approach would train a surrogate model that can predict average spatial temperature variation based on the publicly available data (e.g., outdoor temperature, type of home, weather), and use the predicted average spatial temperature variation as the source of uncertainty of comfort models, capturing spatial and temporal temperature variations without compromising data privacy or requiring extensive instrumentation. Second, the future dynamic model should be able to capture and incorporate the **dynamic patterns** of **HBI** to infer potential lagging factors. For example, dynamic patterns between *TtMSC* and *DoMSC* may indicate the accumulation of small thermal discomforts which may ultimately prompt manual changes. As the dynamic patterns introduced in this manuscript cannot conclusively quantify such lagging factors, future work should strive to capture similar patterns and evaluate the predictive power as potential lagging factors. Third, given that meaningful individual differences are observed from the field data, capturing individual differences across homes and occupants affords opportunities for **personalization** of the comfort model, ultimately delivering greater comfort and optimized energy consumption[40]. Fourth, because DR providers do not know the parameters defining individual differences prior to collecting sufficient thermostat interactions, the future dynamic model should **incorporate changing data by learning over time**, in other words, the model improves as it observes additional real-time data until it captures individual differences. To that end, we believe machine learning techniques designed to continuously update the model based on real-time data, so-called lifelong learning[41] will allow the model itself to continuously evolve for personalization and greater DR capability. The type of dynamic, personalized, continuously learning model of HBI would make possible and comfortable the future promise of GEBs.

**Conclusion**

This paper presents observations and preliminary analysis of occupant thermal comfort and human-building interaction (HBI) data collected during the initial phase of the Whole Energy Homes project. We present observations from the field data including disagreement between field data and ASHRAE 55, considerable spatial temperature variations and their impact on thermal comfort prediction in residential settings, and dynamic behavior for manual set point changes indicating differences across homes and occupants which suggest potentially useful patterns. Based on the observations, we identified critical limitations when applying widely accepted steady-state occupant thermal comfort models to residential Demand Response programs, specifically insufficient prediction performance of discomfort, and insufficient consideration of spatiotemporal temperature variations and individual differences typical of residential settings. Finally, we outline the capabilities required for dynamic comfort models that capture transient and non-uniform indoor conditions common in residential buildings—conditions likely exacerbated by DR events—and incorporate HBI features like time to manual setpoint change (TtMSC) and degree of manual setpoint change (DoMSC) to conform to individual differences. The outcome of this work will guide the development of improved programs for residential grid-interactive energy-efficient buildings that balance the needs of individuals and broader societal goals.

## Acknowledgements

This material is based upon work supported by the U.S. Department of Energy (DOE), Office of Energy Efficiency and Renewable Energy (EERE), Building Technologies Office (BTO) under Award Number DE-EE0009154 and supported by the National Science Foundation (NSF) under Grant No. 2047317. S.K. was partially supported by the Experiential AI Postdoc Fellowship program from Northeastern University and Roux AI Institute. K.B. was supported by the Innovation in Buildings (IBUILD) Graduate Research Fellowship, administered by the Oak Ridge Institute for Science and Education (ORISE) and managed by Oak Ridge National Laboratory (ORNL). ORISE is managed by Oak Ridge Associated Universities (ORAU). All opinions in this paper are the author's and do not necessarily reflect the policies and views of DOE, EERE, BTO, NSF, ORISE, ORNL, or ORAU. The authors also gratefully acknowledge Bethany Sparn and Sugirdhalakshmi Ramaraj from National Renewable Energy Laboratory for their advice.

## Author contributions

S.K., M.K., and D.F. contributed to the conceptualization of the work. S.K., M.K., M. Pavel, and D.F. supervised the research activity and execution. S.K., K.S., M. Pathak, E.C., K.B. contributed to data curation and formal analysis. All authors contributed to the writing and review of the manuscript.

## Data availability statement

The data that support the findings of this study are not yet openly available. At the end of the project, data will be anonymized to protect privacy of participants and will be released publicly. Until the end of the project, controlled access to non-sensitive data is available upon reasonable request to the corresponding author.

## Additional information

**Competing interests** This material is based upon work supported in-kind by ecobee Inc, using their off the shelf thermostats, temperature/occupancy sensors, and window/door sensors products. They did not have significant roles in conceptualization, design, data collection, analysis, decision to publish, or preparation of the manuscript.

# Thermostat Interactions and Human Discomfort: Uncovering Spatiotemporal Variabilities in a Longitudinal Study of Residential Buildings

**SungKu Kang**[1,2+], **Maharshi Pathak**[2+], **Kunind Sharma**[2+], **Emily Casavant**[2], **Katherine Bassett**[2], **Misha Pavel**[3], **David Fannon**[2,4], **Michael Kane**[2,*]

[1] Korea Aerospace University, Department of Mechanical and Aircraft System Engineering, Goyang-si, Gyeonggi-do, 10540, Republic of Korea
[2] Northeastern University, Civil and Environmental Engineering Department, Boston, MA 02115 United States
[3] Northeastern University, Khoury College of Computer Sciences, Boston, MA 02215, United States
[4] Northeastern University, School of Architecture, Boston, MA 02215, United States
* mi.kane@northeastern.edu
+ these authors contributed equally to this work and are listed alphabetically

## Appendix

**Summary of Home Characteristics**

| Home No. | State | City | $A_{floor}$ (ft²) | Home Age (yrs) | Renovation/ Additions | No. of Floors | Years of Occupancy | No. of Occupants | Pre-study Thermostat Type | No. of Thermostats | Equivalent Leakage Area – ELA (ft²) |
|---|---|---|---|---|---|---|---|---|---|---|---|
| 1 | MA | Canton | 2,054 | 72 | 2004 | 2 | 25 | 3 | Programmable | One (1) | 7.8 |
| 2 | MA | Worcester | 2,050 | 36 | N/A | 2 | 13 | 5 | Smart | One (1) | 3.5 |
| 3 | MA | Worcester | 2,051 | 21 | 2016 | 3 | 5 | 5 | Smart | Three (3) | 3.8 |
| 4 | MA | Worcester | 1,612 | 5 | N/A | 2 | 5 | 4 | Smart | One (1) | 1.56 |
| 5 | MA | Worcester | 2,285 | 30-40 | N/A | 2 | 5 | 5 | Programmable | One (1) | 3.1 |
| 6 | MA | Medford | 1,746 | 20 | N/A | 3 | 5 | 4 | Smart | Two (2) | 2.8 |
| 7 | MA | Medford | 2,042 | 122 | 2019 | 2 | <1 | 2 | Programmable | Two (2) | 5.1 |
| 8 | MA | Somerville | 2,882 | 26 | N/A | 3 | 4 | 5 | Programmable | One (1) | 3.8 |
| 9 | MA | Melrose | 2,592 | 120+ | N/A | 4 | 20 | 6 | Manual | One (1) | ---- |
| 10 | MA | Boxford | 6,164 | 40 | N/A | 3 | <1 | 5 | Programmable | Two (2) | 9.9 |
| 11 | CO | Denver | 1,618 | 17 | N/A | 2 | 2 | 4 | Programmable | One (1) | 1.5 |
| 12 | CO | Lafayette | 2,236 | 22 | N/A | 3 | 7 | 4 | Smart | One (1) | 0.68 |
| 13 | CO | Lafayette | 1,371 | 39 | N/A | 2 | 2 | 2 | Programmable | One (1) | 1.21 |
| 14 | CO | Erie | 2,253 | 17 | N/A | 1 | 12 | 2 | Programmable | One (1) | 1.53 |
| 15 | CO | Lafayette | | 29 | N/A | 3 | 15 | 2 | Smart | One (1) | 2.14 |
| 16 | CO | Lafayette | | 36 | N/A | 3 | <1 | 4 | Smart | One (1) | 1.25 |
| 17 | CO | Littleton | 4,072 | 25 | N/A | 3 | <1 | 3 | Smart | One (1) | 2.67 |
| 18 | CO | Broomfield | 2,670 | 13 | N/A | 2 | <1 | 4 | Programmable | Two (2) | 0.78 |
| 19 | CO | Erie | 3,619 | <1 | N/A | 3 | <1 | 5 | Smart | Two (2) | 1.76 |
| 20 | CO | Lafayette | 3,389 | 26 | N/A | 2 | 12 | 5 | Programmable | One (1) | 1.64 |

**Supplementary Table 1:** Summary of participating home characteristics in Whole Energy Home (WEH) research project

| # | T*max | | | T*max > 2 | | | | T*avg | | | T*avg > 2 | | | |
|---|---|---|---|---|---|---|---|---|---|---|---|---|---|---|
| | count | mean | std | count | % | mean | std | count | mean | std | count | % | mean | std |
| 1 | 10.7 | 3.0 | 1.8 | 7.0 | 65.4 | 3.7 | 1.9 | 10.7 | 1.7 | 0.8 | 3.0 | 27.6 | 2.7 | 0.7 |
| 2 | 6.7 | 5.3 | 2.3 | 6.6 | 98.9 | 5.3 | 2.3 | 6.7 | 4.0 | 1.4 | 6.5 | 97.0 | 4.0 | 1.4 |
| 3 | 10.8 | 3.4 | 3.1 | 6.3 | 57.8 | 4.9 | 3.3 | 10.8 | 2.2 | 2.2 | 3.8 | 35.3 | 4.4 | 2.5 |
| 4 | 2.7 | 5.0 | 3.6 | 2.2 | 84.2 | 5.7 | 3.5 | 2.7 | 3.5 | 2.3 | 1.9 | 72.5 | 4.2 | 2.3 |
| 5 | 11.0 | 4.0 | 2.6 | 8.0 | 72.1 | 5.1 | 2.3 | 11.0 | 1.9 | 1.3 | 4.5 | 40.9 | 3.1 | 1.2 |
| 6 | 6.8 | 4.6 | 2.8 | 6.1 | 90.6 | 4.9 | 2.7 | 6.8 | 2.7 | 1.6 | 3.9 | 57.0 | 3.6 | 1.6 |
| 7 | 12.5 | 2.9 | 1.8 | 8.4 | 66.9 | 3.6 | 1.7 | 12.5 | 1.9 | 1.2 | 4.1 | 32.5 | 3.1 | 1.3 |
| 8 | 10.4 | 3.0 | 1.7 | 8.2 | 78.9 | 3.5 | 1.7 | 10.4 | 2.0 | 0.9 | 4.6 | 44.7 | 2.7 | 0.9 |
| 9 | 10.8 | 5.4 | 3.8 | 8.7 | 81.0 | 6.4 | 3.6 | 10.8 | 2.5 | 1.8 | 5.8 | 53.8 | 3.7 | 1.6 |
| 10 | 10.6 | 4.6 | 3.1 | 9.1 | 86.3 | 5.1 | 3.1 | 10.6 | 3.1 | 2.9 | 5.2 | 49.4 | 4.9 | 3.2 |
| 11 | 7.3 | 5.1 | 3.6 | 6.7 | 93.0 | 5.4 | 3.6 | 7.3 | 3.1 | 1.7 | 5.8 | 79.9 | 3.5 | 1.7 |
| 12 | 8.9 | 2.6 | 1.2 | 5.7 | 63.7 | 3.3 | 0.8 | 8.9 | 1.6 | 0.9 | 2.6 | 28.9 | 2.8 | 0.6 |
| 13 | 10.6 | 5.2 | 3.3 | 8.8 | 83.3 | 6.0 | 3.1 | 10.6 | 3.7 | 2.3 | 7.9 | 74.7 | 4.5 | 2.1 |
| 14 | 9.7 | 3.3 | 1.7 | 7.8 | 80.6 | 3.8 | 1.5 | 9.7 | 2.4 | 1.2 | 6.1 | 62.8 | 3.1 | 0.9 |
| 15 | 8.4 | 4.0 | 2.6 | 6.4 | 76.8 | 4.8 | 2.5 | 8.4 | 2.5 | 1.6 | 4.6 | 54.4 | 3.5 | 1.5 |
| 16 | 10.4 | 4.3 | 2.9 | 8.1 | 78.1 | 5.2 | 2.7 | 10.4 | 2.6 | 1.7 | 6.0 | 57.5 | 3.6 | 1.4 |
| 17 | 10.6 | 3.5 | 1.9 | 8.3 | 78.4 | 4.2 | 1.5 | 10.6 | 2.2 | 1.1 | 6.4 | 60.1 | 2.8 | 0.7 |
| 18 | 9.1 | 3.3 | 1.4 | 8.4 | 92.6 | 3.5 | 1.3 | 9.1 | 2.3 | 0.8 | 5.9 | 65.3 | 2.7 | 0.6 |
| 19 | 9.5 | 5.0 | 2.9 | 8.5 | 89.4 | 5.4 | 2.8 | 9.5 | 2.4 | 1.3 | 5.5 | 58.3 | 3.2 | 1.1 |
| 20 | 9.5 | 6.0 | 2.2 | 9.3 | 98.0 | 6.1 | 2.1 | 9.5 | 4.2 | 1.8 | 8.5 | 90.2 | 4.5 | 1.7 |

**Supplementary Table 2.** Detailed statistics of spatial temperature variations across the home (**Figure 5**). Count is the thousands count of time series temperature samples. Mean (°F), standard deviation (°F), and percentage of time (%) that spatial temperature variations exceed 2°F. These statistics are calculated for T*max, the absolute difference at each sample time between the thermostat temperature reading and the remote sensor in an occupied room with the greatest temperature difference from the thermostat, and T*avg, the average of absolute difference between the thermostat temperature reading and each occupied room temperature at each sample time.

| Percentile | A (min. / °F) | B (min.) | RMSE |
|---|---|---|---|
| 50 | 1.269 | 54.22 | 7.18 |

**Supplementary Table 3. Coefficients and Root Mean Square Error (RMSE) of the Linear Regression Curves Representing the 50th Percentile of MSCs.** The table below provides the coefficients and RMSE values for the linear regression curves corresponding to the 50th percentile of manual setpoint changes (MSCs). These curves represent the relationship between the degree of manual setpoint change (DoMSC) and the time to manual setpoint change (TtMSC) within 150 minutes, as shown in Figure 11. The linear fitting function is expressed as $y = Ax + B$, where $A$ represents the slope, and $B$ represents the intercept.